\documentclass[twocolumn]{article}
\usepackage[top=2cm, bottom=2cm, left=1.5cm, right=1.5cm]{geometry}
\usepackage{setspace}
\usepackage{float}
\usepackage{amsmath}
\usepackage{graphicx}
\usepackage{amssymb}
\usepackage{color}

\begin{document}
\twocolumn[
\title{Temperature Integration: an efficient procedure
for  calculation of free energy differences}

%\author{Farhi et al.}
%%\author{\small{Asaf Farhi$^1$, Guy Hed$^{1,2}$, Michael Bon$^3$, Nestor Caticha$^4$, Amir Marcovitz$^5$, Chi H Mak$^6$ and Eytan Domany$^1$}\\
\author{\small{Asaf Farhi$^{1,\dagger}$, Guy Hed$^{1,2}$, Michael Bon$^3$, Nestor Caticha$^4$, Chi H Mak$^5$ and Eytan Domany$^1$}\\
\\
\small{$^1$ Physics Department Weizmann Institute of Science, Rehovot 76100, Israel}\\
\small{$^2$ Center for Educational Technology, Tel Aviv 61394, Israel}\\
\small{$^3$ Institut de Physique Th\'{e}orique, CEA Saclay, CNRS, 91191 Gif-sur-Yvette, France}\\
\small{$^4$ Institute of Physics, University of S\~{a}o Paulo, S\~{a}o Paulo, SP, Brazil}\\
%\small{$^5$ Department of Structural Biology Weizmann Institute of Science, Rehovot 76100, Israel}\\
\small{$^5$ Department of Chemistry University of Southern California, CA 90089-0482, USA}\\
}

\date{\today}

\maketitle

\abstract{ We propose a method, Temperature Integration, which
allows an efficient calculation of free energy differences between
two systems of interest, with the same degrees of freedom,
which may have rough energy landscapes. The
method is based on calculating, for each single system, the difference
between the values of $\ln Z$ at two temperatures, using a Parallel
Tempering procedure. If our two systems of interest have the
same phase space volume, they have the same values of $\ln Z$ at
high-$T$, and we can obtain the free energy difference between
them, using the two single-system calculations described above.
If the phase space volume of a system is known, our method can
be used to calculate its absolute (versus relative) free energy
as well. We apply our method and demonstrate its efficiency on
 a ``toy model'' of hard
rods on a 1-dimensional ring. }
\medskip]
\section{Introduction}
\let\thefootnote\relax\footnote{$^{\dagger}$asaf.farhi@gmail.com}

Calculating free energy differences between two physical systems,
or between two thermodynamic states of the same system, is a topic
of considerable current interest. The problem arises mainly in soft condensed
matter, especially in studies of macromolecules such as proteins or
RNA. When the systems in question
have complex energy landscapes with many local minima, generating
an equilibrium ensemble of configurations in reasonable running time
becomes a major challenge for computational physics. Indeed, a variety
of advanced methods and algorithms have been introduced to answer
the challenge, both in the context of Molecular Dynamics and Monte
Carlo (for recent reviews see \cite{chipot2007free,Zuckerman_Equilibrium_sampling,frenkel1996understanding,shirts2007alchemical,binder2010monte}).

Free energy difference between two systems can be calculated using equilibrium methods (as used by us) and non equilibrium methods.
Existing equilibrium methods are composed of 3 stages: (1) selection of intermediates that interpolate between the systems
(2) ergodic sampling of the system at each intermediate and (3) calculation of the free energy difference between the
 systems using one of the methods mentioned below. The commonly used methods include Bennett Acceptance Ratio
  \cite{bennett1976efficient}, Weighted Histogram Analysis Method \cite{kumar1992weighted} Exponential
 Averaging / Free Energy Perturbation \cite{zwanzig1954high} and Thermodynamic Integration (ThI)
\cite{frenkel1996understanding,kirkwood1935statistical,straatsma1991multiconfiguration}.

Non equilibrium methods measure the work needed in the process of
switching between the two Hamiltonians. These methods
use Jarzynski relations \cite{jarzynski1997nonequilibrium}
("fast growth" is one of their variants \cite{hendrix2001fast})
and its subsequent generalization by Crooks \cite{crooks1999path}.

Free energy differences are calculated in several contexts, including binding free energies \cite{kollman1993free,woods2011water,deng2009computations}
, free energies of hydration \cite{straatsma1988free}, free energies of solvation
\cite{khavrutskii2010computing} and of transfer of a molecule from gas to solvent \cite{shirts2007alchemical}.
Binding free energy calculations are of high importance since they can be used for molecular docking
\cite{meng2011molecular} and have potential to play a role in drug discovery \cite{chodera2011alchemical}.

Most of the applications mentioned above can be tackled from a different direction using methods which measure the
free energy as a function of a reaction coordinate. These methods include Adaptive Biasing Force
\cite{darve2001calculating} and Potential Mean Force \cite{kirkwood1935statistical}.

Our novel method, Temperature Integration (TempI) can, in principle, be used instead of equilibrium methods.
In order to demonstrate the advantages of TempI, we chose to introduce the idea in the context
of Thermodynamic Integration
(ThI) \cite{frenkel1996understanding,kirkwood1935statistical,straatsma1991multiconfiguration}.
ThI is based on simulating a set of systems defined by different
values of a parameter $0 \leq \lambda \leq 1$, where the two systems that we wish
to compare are realized when $\lambda = 0$ or 1. The free energy difference is given
as an integral over $\lambda$, which is evaluated numerically. Hence $L$, the number of
values of $\lambda$ one needs, depends on how fast the integrand varies, which in turn is
determined by the dissimilarity of the two compared systems. In general, the optimal
choice of the intermediate systems is a challenge \cite{chodera2011alchemical}.

Since in many cases of interest each of the systems studied
has a complex energy landscape with minima separated by large barriers,
equilibration times are long. A favored choice to alleviate this problem is Parallel Tempering (PT) e.g.
\cite{lin2003parallel} or replica exchange method \cite{ferrenberg1988new}\cite{Zuckerman_Equilibrium_sampling} in the context of MD (Hamiltonian Replica Exchange is a variant in the $\lambda$ dimension \cite{fukunishi2002hamiltonian}). This technique necessitates equilibration
of a system of $N$ particles at a set of $n\sim\sqrt{N}$ inverse temperatures $\beta_{k},k=1,...n$ (where $N$ is the number of particles).

Combination of ThI and PT has been suggested by others \cite{woods2003development}\cite{rodinger2005enhancing}
as an efficient way \cite{sugita2000multidimensional} to calculate
free energies of such systems. Since simulations of $n$
replicas of the system are performed at each of $L$ values of $\lambda$, using PT with ThI calls for simulations
at a set of $L \times n$ points in the $\lambda,T$ plane (see Figure \ref{fig:grid}).

Our novel method, TempI, uses the
temperature dimension, explored by parallel tempering, for the
calculation of free energy differences; In effect the replicas, simulated
in the parallel tempering procedure, are used as intermediates for
the calculation of free energy differences. Thus, the need for sampling both $T$ and $\lambda$ dimensions is eliminated.
%Hence clearly the challenge of choosing
%appropriate intermediates
%\cite{chodera2011alchemical} also disappears.
Furthermore, since in TempI the internal energy
$\left\langle H\right\rangle$ is a monotonic function of $\beta$,
the choice of intermediates is no longer a challenging problem\cite{chodera2011alchemical}
(see Appendix for details), and the calculation is much easier to
verify.
%Moreover, the monotonicity of the function may result in needing
%less intermediates and facilitates the calculation (the number of
%integration points scales as the free energy difference which is
%roughly linear with the number of particles $N$).

Temperature Integration is based on calculating, for each system,
the difference between $\ln Z$ at the temperature of interest and
at a high temperature, using parallel tempering procedures. In
case the two compared systems have the same phase space and hence  $Z$
at high-$T$, the difference between $\ln Z$ of the systems
at the temperature of interest can be calculated and the free
energy difference is obtained. For cases when the two compared systems have
different $Z$ value at high-$T$, we use an additional
methodological advance that enables the comparison. For some systems
the calculation of absolute free energy values, using
Temperature Integration, is also feasible.

The structure of the paper is as follows. In Section
\ref{sub:The-method} we introduce the method of Temperature
Integration. In Section \ref{sec:cutoff} we describe how to
compare two systems with different values of partition functions
at high temperatures. In Section \ref{sec:toy} we apply the method
to a toy problem, demonstrate a calculation of absolute free
energy values, and compare its performance with that of ThI combined
with PT. The work is summarized in Section \ref{sec:Discussion}.

%
%The method is used for the physical problem of interior loops of
%RNA, and its efficiency is demonstrated for a toy problem of interacting
%particles on a ring.
%
%The structure of the paper is as follows. In Section \ref{sec:The-physical-problem}
%the problem of the interior loop of RNA is introduced and described in detail
%and the ThI and PT methods are reviewed briefly. In Section \ref{sub:The-method}
%we introduce Temperature Integration and describe a few additional
%methodological advances. In section \ref{sec:loop} we describe in detail the
%RNA simulation, and present its results. In Section \ref{sec:toy} we
%compare the efficiency of TempI relative to ThI with PT, using a simple toy model.
%The work is summarized in Section \ref{sec:Discussion}.
-----------------------------
\section{\label{sub:The-method}Calculation of $\Delta F$ by Temperature
Integration}

Consider two systems, denoted by $A$ and $B$, at a given temperature $T_1 =
\beta_1^{-1}$, between which we want to calculate the free energy
difference $\Delta F_{A\rightarrow{B}}\left(\beta_{1}\right)$.
We will first assume that the two systems have the same degrees of freedom and
phase space volume -
so as $\beta\to0$ they have the same value of the partition function
(the assumption of having the same value of the partition function will be relaxed in section \ref{sec:cutoff}).
One of the most commonly used methods
for calculating such free energy difference between two such systems is Thermodynamic Integration \cite{frenkel1996understanding,kirkwood1935statistical,straatsma1991multiconfiguration}, which we now briefly describe.

\subsection{\label{sub:Thermodynamic-integration}Thermodynamic Integration}

%This method is used to calculate the free energy difference between
%two systems with the same degrees of freedom, by varying a parameter
%$\lambda$ that interpolates between the two compared systems. We
%will use the systems $\tilde{C}$ and $\tilde{G}$ described above
%to demonstrate application of this general method in a specific context.

Denote the Hamiltonians of the two systems by $H_{A}(\mathbf{\boldsymbol{\Omega}})$
and $H_{B}(\mathbf{\boldsymbol{\Omega}})$, where $\Omega$
denotes the coordinates of the system. Noting that the two systems
have the same coordinate space, we define a $\lambda$-weighted
hybrid system, characterized by the Hamiltonian $H(\lambda)$:
\begin{equation}
H(\lambda,\mathbf{\boldsymbol{\Omega}})=\lambda H_{B}(\mathbf{\boldsymbol{\Omega}})+(1-\lambda)H_{A}(\mathbf{\boldsymbol{\Omega}})\label{eq:Hamiltonian}\end{equation}
As shown in \cite{frenkel1996understanding,straatsma1991multiconfiguration},
the free energy difference is given by \begin{equation}
\triangle F_{A\rightarrow B}\left(\beta_{1}\right)=\intop_{0}^{1}\left\langle \frac{dH}{d\lambda}\right\rangle d\lambda=
 \end{equation}
 \begin{equation}
\intop_{0}^{1}\left[\left\langle H_{B}(\mathbf{\boldsymbol{\Omega}})\right\rangle _{\lambda}-\left\langle H_{A}(\mathbf{\boldsymbol{\Omega}})\right\rangle _{\lambda}\right]d\lambda\label{eq:reg} \end{equation}
 where $\left\langle X\right\rangle _{\lambda}$ denotes the equilibrium
average of $X$ in the ensemble characterized by $H(\lambda)$. The
expression for $\triangle F_{A \rightarrow B}\left(\beta_{1}\right)$,
written explicitly, takes the form:
\begin{equation}
\triangle F_{A\rightarrow B}\left(\beta_{1}\right)=
\end{equation}
 \begin{equation}
\intop_{0}^{1} \int\frac{\left[H_{B}(\mathbf{\boldsymbol{\Omega}})-H_{A}(\mathbf{\boldsymbol{\Omega}})\right]e^{-\beta_{1}\left[\lambda H_{B}(\mathbf{\boldsymbol{\Omega}})+\left(1-\lambda\right)H_{A}(\mathbf{\boldsymbol{\Omega}})\right]}d\mathbf{\boldsymbol{\Omega}}}{Z(\lambda)} d\lambda\label{eq:detailed}\end{equation}
 with \begin{equation}
Z(\lambda)=\int e^{-\beta_{1}\left[\lambda H_{B}(\mathbf{\boldsymbol{\Omega}})+\left(1-\lambda\right)H_{A}(\mathbf{\boldsymbol{\Omega}})\right]}d\mathbf{\Omega}\end{equation}
The integration is performed numerically, with the integrand evaluated
at each one of a set of values of $\lambda$ by Monte Carlo simulations.
As implied by \eqref{eq:detailed}, the two systems are required to
have the same degrees of freedom. The complexity of ThI is proportional
to the number of $\lambda$ values $L$, required to estimate the
integral within a given error range. In the best case scenario of
$\left\langle H_{ B} - H_{A}\right\rangle_\lambda$ being a monotonic function of $\lambda$, this number
increases roughly linearly with $\Delta F$. Hence $L$ is roughly proportional
to the number of particles: $L \sim\Delta F\sim N$.

\subsection{\label{sub:TI-and-parallel-tempering}Thermodynamic Integration with
Parallel Tempering}

In many cases of interest both systems $A$ and $B$, and hence also all the $\lambda$-weighted
intermediate systems, have rugged energy landscapes
with many local minima. As the decorrelation times grow exponentially
with $\triangle E/k_{\rm B}T$ , where $\triangle E$ is the energy barrier
between nearby valleys, equilibration times in these systems can be
rather long. In order to overcome this problem, and obtain the equilibrated thermodynamic
averages $\left\langle H(\Omega)\right\rangle _{\lambda}$, one can
use the Parallel Tempering procedure \cite{hansmann1997parallel,earl2005parallel}.
However, implementing both thermodynamic integration and parallel
tempering yields an unnecessary overhead in running time, as we will
demonstrate.

In order to implement thermodynamic integration we first choose a
set of values $\lambda_{i},~i=1,...L$, that will enable us to have
a good sampling of the function $\left\langle H(\mathbf{\boldsymbol{\Omega}})\right\rangle _{\lambda}$
for the integration in equation \eqref{eq:reg}. Thus, $L$ is related to
the desired precision of the integration.

In principle parallel tempering should be performed for each $\lambda_{i}$,
simulating each of the $m$ $\lambda_{i}$-weighted systems over a
set of temperatures, given by $\beta_{k},k=1,...n\sim\sqrt{N}$ \cite{hansmann1997parallel}.
Finally, using the calculated values of $\left\langle H(\mathbf{\boldsymbol{\Omega}})\right\rangle _{\lambda_{i}}$
at a temperature of interest $\beta_{1}$, we approximate the integral
of Equation \eqref{eq:reg} by a sum over $L$ terms, to get the free energy
difference $\triangle F_{A \rightarrow B}\left(\beta_{1}\right)$.

\begin{figure}[h]
\begin{centering}
\includegraphics[width=8cm]{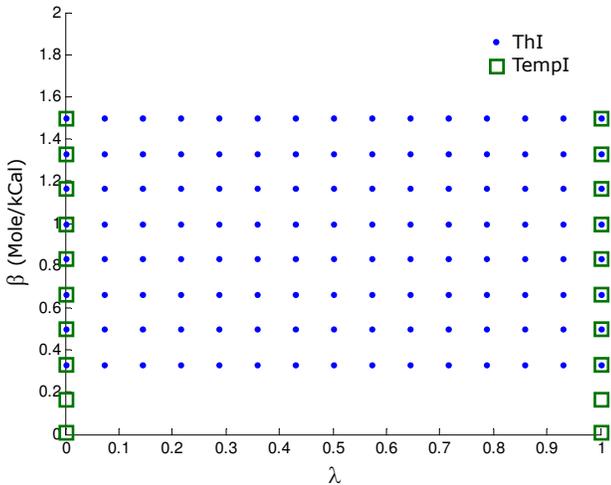}
\caption{\label{fig:grid}Illustration of the grid of values over which the
Monte-Carlo simulations are performed in the two procedures:
ThI with PT and TempI. The lowest values of $\beta$
correspond to some very high finite temperature. }
\end{centering}
\end{figure}

The procedure involves running Monte-Carlo simulations over $L$ $\lambda$-values,
and $n$ $\beta$-values, that is, over a grid of $L \times n$ instances
of the hybrid system. An illustration of this
grid is presented in figure \ref{fig:grid}.

\subsection{\label{sub:The-method}Calculation of $\Delta F$ by Temperature
Integration}

%\subsection{The method}

We present now our method, which obtains the free energy difference $\triangle F_{A \rightarrow B}\left(\beta_{1}\right)$,
using only simulations that
are done in the
process of parallel tempering, performed for the two systems $A$ and $B$,(eliminating the need for simulations at
a set of $\lambda_{i}$ values).
This method can be applied to any two systems that have the same degrees
of freedom $\Omega$.

As $\beta\to0$, the limiting value of the partition function of a
system yields the phase space volume. In particular, if systems $A$
and $B$, which have the same coordinate space $\boldsymbol{\left\{ \Omega\right\} }$,
have the same $\beta\to0$ limit, we have \begin{equation}
Z_{B}\left(\beta\to0\right)=Z_{A}\left(\beta\to0\right)=\int d\boldsymbol{\Omega}\,.\label{eq:Z0Z0}\end{equation}

In this case we can use the following identity to obtain, for a
finite $\beta_{1}$, the difference of the free energies:

\begin{equation}
\ln Z\left(\beta_{1}\right)-\ln Z\left(\beta\to0\right)=\int_{0}^{\beta_{1}}\frac{d\ln Z}{d\beta}d\beta=-\int_{0}^{\beta_{1}}\left\langle H\right\rangle d\beta\,\label{eq:intH}\end{equation}
Using equations (\ref{eq:Z0Z0}) and (\ref{eq:intH}) we obtain\begin{eqnarray}
\triangle F_{A \rightarrow B}\left(\beta_{1}\right) & = & \frac{1}{\beta_{1}}\left[\ln Z_{A}\left(\beta_{1}\right)-\ln Z_{B}\left(\beta_{1}\right)\right]\nonumber \\
 & = & \frac{1}{\beta_{1}}\left[\intop_{0}^{\beta_{1}}\left\langle H_{B}\right\rangle d\beta-\intop_{0}^{\beta_{1}}\left\langle H_{A}\right\rangle d\beta\right]\,.\label{eq:DFCG}\end{eqnarray}
For each of the systems $A$ and $B$
we estimate the integrals on the right hand side by procedures of
parallel tempering, sampling the system at a series of values $\beta_{1},\ldots,\beta_n$.
We choose values such that the highest temperature sampled (corresponding
to $\beta_{n}$) is much larger than the internal energy of the system
at $\beta_{1}$, so $Z(\beta_{n})\simeq Z(\beta\to0)$ is satisfied.

\section{Comparing two systems with different partition functions at high Temperatures by introducing a cutoff\label{sec:cutoff}}
The condition of Equation \eqref{eq:Z0Z0} poses a problem for any two systems that have different partition functions
 values at high -$T$. In order to satisfy the condition stated above we had to set a cutoff over the interactions,
$E_{\mathrm{cutoff}}$. We show that our results do not depend on
the choice of $E_{\mathrm{cutoff}}$ and $\beta_{n}$, as long as
$E_{\mathrm{cutoff}}$ is much larger than any typical interaction
energy in the system at $\beta_{1}$, and $\beta_{n}^{-1}\gg E_{\mathrm{cutoff}}$.

Note that this use of cutoff is general and can
be used for any energy term that differentiates between the systems
at the high $T$ limit.

The proposed calculation of the free energy difference between the
two systems at the temperature of interest $\beta_{1}$ is legitimate
only if our choice of the cutoff energy has a negligible effect on
the partition function value of each of the two systems at $\beta_{1}$.
In addition, the highest temperature used, corresponding to $\beta_{n}$,
must be such that the equality of the partition functions of the two systems is
satisfied to a good accuracy.

We denote the Hamiltonian with the cutoff energy by $H'$, and write
the requirements stated above explicitly as follows:
\begin{equation}
lnZ_{B}\left(\beta_{1},H\right)\simeq lnZ_{B}\left(\beta_{1},H'\right)\label{eq:cond1}\end{equation}
\begin{equation}
lnZ_{A}\left(\beta_{1},H\right)\simeq lnZ_{A}\left(\beta_{1},H'\right)\label{eq:cond2}\end{equation}
\begin{equation}
lnZ_{B}\left(\beta_{n},H'\right)\simeq lnZ_{A}\left(\beta_{n},H'\right)\label{eq:cond3}\end{equation}
In order for the cutoff to have a negligible effect on the partition
functions at the temperature of interest it has to be set to a value
that satisfies \begin{equation}
E_{\mathrm{cutoff}}\gg k_\mathrm{B}T_{1}.\label{eq:cond4}\end{equation}

As for $\beta_{n}$, if the cutoff energy satisfies \begin{equation}
E_{\mathrm{cutoff}}\ll k_\mathrm{B}T_{n},\label{eq:cond5}\end{equation}
 the systems will have almost equal probability to be in all the regions
of their phase space, including ones which were restricted due to high energy values. Thus, the partition functions values
 of the two systems will be almost equal.

Hence if these requirements are satisfied one can write:
\begin{eqnarray}
&lnZ_{A}\left(\beta_{1},H\right)-lnZ_{B}\left(\beta_{1},H\right)\simeq&\notag\\
&lnZ_{A}\left(\beta_{1},H'\right)-lnZ_{B}\left(\beta_{1},H'\right)\simeq&\notag\\
&lnZ_{B}\left(\beta_{n},H'\right)-lnZ_{B}\left(\beta_{1},H'\right)&\notag\\
&-\left[lnZ_{A}\left(\beta_{n},H'\right)-lnZ_{A}\left(\beta_{1},H'\right)\right]&
\end{eqnarray}
Using the identity in Eq. \eqref{eq:intH}, we can write:
\begin{eqnarray}
&\triangle F_{A\rightarrow {B}}\left(\beta_{1},H\right)=\frac{1}{\beta_{1}}\left[lnZ_{A}\left(\beta_{1},H\right)-lnZ_{B}\left(\beta_{1},H\right)\right]\simeq&\notag\\
&\frac{1}{\beta_{1}}\left[\intop_{\beta_{n}}^{\beta_{1}}\left\langle H'_{B}\right\rangle d\tilde{\beta}-\intop_{\beta_{n}}^{\beta_{1}}\left\langle H'_{A}\right\rangle d\tilde{\beta}\right]&
\label{eq:deltaFCtoG}
\end{eqnarray}
So the calculation of free energy will be negligibly affected by the
use of a cutoff energy as long as we fulfill the relevant conditions.

This use of cutoff energy is relevant also to ThI and similar methods.
Consider for example two systems, $H_0$ and $H_1$ that have different steric constraints (resulting
in different $\beta\to0$ limits), Then e.g. at $\lambda=0$ there may be micro-states
with finite statistical weight $e^{-\beta H_0}$ and infinite
energy in $H_1$, so the sampling of the internal energy
$\langle H_1\rangle_{\lambda=0}$
is infeasible.

It can be seen that since the partition function of a system
that has the Hamiltonian with the cutoff $H'$ is almost equal to
the one of a system with the Hamiltonian $H$ at $\beta_1$. Thus
\begin{equation}
\triangle F_{A\rightarrow {B}}\left(\beta_{1},H\right)\simeq
\frac{1}{\beta_{1}}\left[lnZ_{A}\left(\beta_{1},H'\right)-
lnZ_{B}\left(\beta_{1},H'\right)\right],
\end{equation}
and ThI can be implemented for systems with $H'$ and yield almost the same
result.

In conclusion, with the use of cutoff energy in ThI, that enables
us to sample the integrand in all cases, the free energy difference
between any two systems that have the same degrees of freedom can
be calculated.

\section{Applying Temperature Integration to a toy model\label{sec:toy}}

In the following sections we present the results obtained by applying
Temperature Integration with interaction cutoff
to a ``toy model'', of $N=8-25$ particles on a ring (one dimension,
periodic boundary conditions) in an external potential, interacting
via a hard core potential. For this model we evaluated
the free energy difference between two systems, which differ in the
size of their hard cores. We did this in two ways: First we performed
ThI at a set of $\lambda$ values, where at each value the corresponding
system was equilibrated using PT; second, we used Temperature Integration.
We compared the results as well as the amount of (computational) work needed,
for the two ways, to achieve similar accuracy. We estimated the gain
in work (number of Monte Carlo steps needed) and the way it scales
with the number of particles.

\subsection{Definition of the model}

Here we demonstrate the method for a toy model in one dimension with periodic boundary conditions:
we place $N$ particles on the unit circle, with the position
of a particle defined by an angle ($\theta=\left[0,2\pi\right]$).
The particles are in an external potential, given by
\begin{equation}
V\left(\theta_{i}\right)=V_{0}\cdot cos\left(2\theta_{i}\right)
\end{equation}
and have ``hard core'' interaction:
\begin{equation}
U\left(\theta_{i},\theta_{j}\right)=\left\{ \begin{array}{cc}
\infty & dist\left(\theta_{i},\theta_{j}\right)<W\\
0 & dist\left(\theta_{i},\theta_{j}\right)\geqslant W\end{array}\right.
\end{equation}
Here $W$ is the size of a particle's ``core'',
and $dist\left(\theta_{i},\theta_{j}\right)$ is the angle difference
between the particles.

In order to apply TeI we introduced a cutoff of this potential, replacing
$U\left(\theta_{i},\theta_{j}\right)$ by:

\begin{equation}
U\left(\theta_{i},\theta_{j}\right)=\left\{ \begin{array}{cc}
U_\mathrm{cutoff} & dist\left(\theta_{i},\theta_{j}\right)<W\\
0 & dist\left(\theta_{i},\theta_{j}\right)\geqslant W\end{array}\right.\end{equation}

where $U_\mathrm{cutoff}$ is the energetic cost of two particles having overlapping
cores. The effect of the cutoff on the results of the calculation
is negligible as shown below in \ref{sub:Absolute-Values-of}.

We calculate the free energy difference between two systems $A$ and
$B$, that have the same number of particles and different
values of  $W$. Specifically, we set the constants $V_{0},\, W$
in the systems $A,B$ to have the following values\footnote{
The $\frac{1}{N}$ factor was added to maintain constant density
of particles.}:
\begin{equation}
V_{0A}=V_{0B}=4,\qquad W_{A}=\frac{\pi}{4}\frac{1}{N},\qquad W_{B}=\frac{\pi}{8}\frac{1}{N}\end{equation}
We work at temperature $k_\mathrm{B}T_{1}=0.6157$ so that the barrier associated
with the external potential is significantly higher than $k_{\rm B}T_{1}$,
and set $U_\mathrm{cutoff}=70$, so that $U_\mathrm{cutoff}\gg k_\mathrm{B}T_{1}$ and hence condition
\eqref{eq:cond4} is satisfied.

\subsection{Details of the Monte Carlo Simulation}

In the initial configuration all the particles were placed in the
interval $\left[0,\pi\right]$, at equal distances.

The local Monte Carlo move consists of randomly choosing one of the
particles and changing its position $\theta$ to $\theta+\Delta\theta$,
where $\Delta\theta \in \left[-\frac{\pi}{128},\frac{\pi}{128}\right]$
is selected with uniform probability.

All the Monte Carlo simulations are performed using Parallel Tempering;
that is, after a certain number of local moves (500 in our case),
we attempt to exchange configurations between systems at adjacent
temperatures. We simulated systems with 8,10,12,14,17,20 and 25 particles
and performed up to $\left(1,1.1,1.2,1.35,1.5,2\right)\cdot10^{8}$
MC moves in total for each one respectively.
%(for a given temperature).

\subsection{Using Thermodynamic Integration and PT\label{sub:Thermodynamic-Integration-Toy}}

In order to compare the results obtained by TeI we used ThI in combination with PT.
The set of temperatures at which we worked was
selected as follows.

The highest temperature was chosen as $3V_{0}/k_\mathrm{B}$, to
enable particles to cross easily the energy barrier ($2V_{0}$) of
the external potential.

The temperatures for both systems $A$ and $B$ were first selected
so that the acceptance rates for exchanging configurations at neighboring
temperatures will be between $0.25$ and $0.35$. Then, the set of
temperatures to be used was of that system in which the product of
the acceptance rates was lower\textbf{ }(i.e using the set of temperatures
with higher density).

The integral in \eqref{eq:deltaFCtoG} was calculated by adding sampling
points according to global adaptive Simpson's quadrature (GASQ) \cite{lyness1969notes,malcolm1975local}.
That is, for each value of $\lambda$, a simulation of a compound system was performed at the selected set of
temperatures, and $\left\langle H_{B}(\mathbf{\boldsymbol{\Omega}})\right\rangle _{\lambda,\beta_{1}}-\left\langle H_{A}(\mathbf{\boldsymbol{\Omega}})\right\rangle _{\lambda,\beta_{1}}$
was calculated.

At each division of an interval, we sampled the internal energy at
another 4 $\lambda$ values (according to GASQ \cite{lyness1969notes,malcolm1975local}),
calculated the integral with the current set of intervals and displayed
the result as a function of the total number of MC steps performed.

\subsection{Using Temperature Integration\label{sub:Temperatures-Integration}}

For each system, we evaluated numerically the integral:
\begin{equation}
k_\mathrm{B}T_{1}\intop_{\beta_n}^{\beta_{1}}\left\langle H\right\rangle d\beta\label{eq:TempIint}\end{equation}
 We chose $k_\mathrm{B}T_n$ to be given by $k_\mathrm{B}T_n=400*U_{0}*N*(N-1)/2$ in
order to satisfy condition (\ref{eq:cond5}).

This was done iteratively.
In each iteration we performed parallel tempering on a different set
of temperatures and calculated the internal energies.
Then, we added the results to those of the former iteration and calculated
the integral (\ref{eq:TempIint}) numerically according to the current
set of sampling points ($\beta$-values and the corresponding internal
energies).

After each calculation of the integral, we registered the result as
well as the number of Monte Carlo steps performed in total until a
stop criterion was reached (see Appendix).

The method by which we chose the sets of $\beta$-values was based on the
global adaptive Simpson's quadrature and was suited to maintain optimal
acceptance rates between the systems (as explained in the Appendix).

\subsection{Absolute Values of free energy and Verification of the method \label{sub:Absolute-Values-of}}

The highest temperature $T_n$ was chosen to be much higher than
the total interaction energy, so the
partition function becomes the phase space volume, $\Omega$.
Hence, when the phase space volume of a system is known at $T_{n}$,
we can calculate the absolute (not relative) free energy of the
system:
\begin{equation}
F=-k_\mathrm{B}T_{1}\left[\ln\frac{\Omega}{l^{3N}}+\int_{\beta_{n}}^{\beta_{1}}\left\langle H\right\rangle d\beta\right]\;,
\label{eq:absolute}\end{equation}
where the integral in equation \eqref{eq:absolute} can be estimated using
the methods described in Section \ref{sub:The-method}.

In many systems (including our toy model) when we neglect all interactions
(including steric) the phase space volume is known and the values
of free energy can be calculated according to \eqref{eq:absolute},
enabling the comparison of free energies of systems with different
degrees of freedom.

To assess the accuracy of TempI (also in the context of absolute free energy values) for our toy
model, we compared its result (for $N=3$ particles) with ``exact enumeration'' numerical evaluation of the free energy,
 obtained by performing with high precision the integral
\begin{equation}
F=-k_\mathrm{B}T_1 \ln\left(\frac{\int e^{-\beta_{1}H}d\mathbf{\Omega}}{l^{3N}}\right)
\end{equation}
In this integration, the microstates in which the steric limitations
were violated weren't taken into account, enabling us a liable comparison
of the results according to \eqref{eq:absolute}.

\subsection{Results\label{sub:Results}}

\begin{figure}[h]
\begin{centering}
\includegraphics[width=8cm]{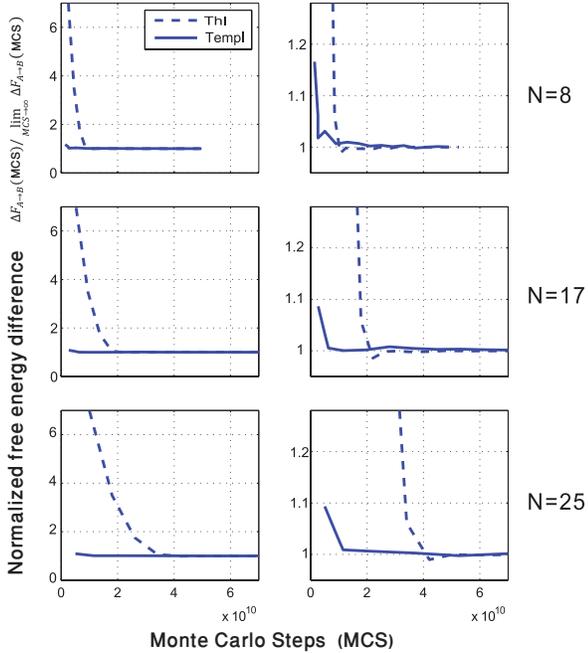}
\par\end{centering}
 \caption{\label{fig:Free-energy-difference}The result for $\Delta F$, normalized by the
``asymptotic'' value, as a function of the number MC steps
 using ThI and TempI, for systems with 8,17,25 particles.}
\end{figure}

%##
ThI (with PT) and TempI converged to the same
asymptotic value of the $\Delta F_{A\to B}$ after a large enough number MC steps.
The number of MC steps required for each method to approach the asymptotic value is,
however, very different. In Fig. \ref{fig:Free-energy-difference},
we present the relation between the values of $\Delta F_{A\to B}$, calculated
by the two methods, and its asymptotic value, as a function of the \textbf{total} number of
the MC steps used in the calculation, for systems with 8, 17 and 25 particles.
Evidently, in
the TempI procedure the convergence of $\Delta F_{A\to B}$ to the asymptotic value
was significantly faster than in ThI.

\begin{figure}[h]
\begin{centering}
\includegraphics[width=6cm]{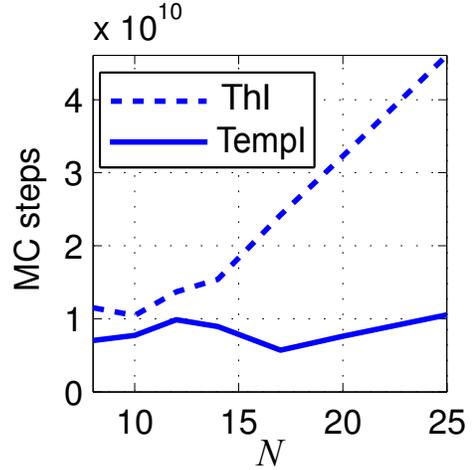}
\par\end{centering}
\caption{\label{fig:MC-steps-needed-as-function-of-N}MC steps needed for $1\%$
convergence in both methods as a function of the number of particles, $N$.}
\end{figure}

Another measure of the relative efficiencies of the two methods is
the number of MC steps needed to achieve a desired level of accuracy.
In Fig. \ref{fig:MC-steps-needed-as-function-of-N} we present the
number of MC steps needed to approach the asymptotic value to within
$1\%$, by both methods, as a function of the number of particles.
The ratio between the number of MC steps needed for convergence
in ThI and in TempI increases steeply with particle number, as
demonstrated in Fig. \ref{fig:MC-steps-needed-as-function-of-N}.
We also computed the absolute free energies for $N=3$ particles,
by both TempI and by exact enumeration.
The numerical results obtained by TempI, using \eqref{eq:absolute}, were:
\begin{eqnarray}
F_{A}&=-k_\mathrm{B}T_{1}\left[ln\Omega\left(T_{n}\right)+\int_{\beta_{n}}^{\beta_{1}}
\left\langle H\right\rangle d\beta\right]&\\
&=-k_\mathrm{B}T_{1}\left[5.51+12.72\right]=-11.23&\\
F_{B}&=-k_\mathrm{B}T_{1}\left[5.51+13.40\right]=-11.64\;,&
\end{eqnarray}
where the phase space volume is $\Omega=\left(2\pi\right)^{3}$.
Numerical integration over the whole phase space
yielded the same values, confirming the validity of our method in general and in the context of absolute free energy
values.

\section{Discussion\label{sec:Discussion}}

We presented Temperature Integration, a method to calculate the free
energy difference between two systems at some inverse temperature
$\beta_{1}$. Temperature Integration is an efficient method since
the temperatures used in the parallel tempering procedure are used
as intermediates in the calculation of free energy. Moreover, the
method for choosing the intermediates for the calculation of free
energy in TempI is general (see Appendix for details). Hence, the
method is robust, which is important for automation and
high-throughput use.

In the calculation, we performed parallel tempering
procedures for each of the two systems over sets of temperatures between
$\beta_{1}$ and $\beta_{n}$, where $\beta_{n}^{-1}\gg E_{\mathrm{cutoff}}$
- the maximal interaction energy of the system. Then, we use the internal
energy values obtained at each temperature to estimate numerically
the two integrals in equation \eqref{eq:DFCG}, and hence the free
energy difference between the two systems.

Furthermore, absolute values of the free energy can be calculated
for systems in which the phase space volume is known (when all
interactions are neglected) as stated in section
\ref{sub:Absolute-Values-of}.

TempI can be used also for systems with a smooth energy landscape (where PT is not needed).
TempI has the advantage of simplicity (saves programming time) since the simulations are performed only on
the two original systems (i.e. using only the two "pure" Hamiltonians).

While in other methods for calculating free energy difference
between two systems the choice of appropriate intermediates
remains a challenge \cite{chodera2011alchemical}, in TempI the
internal energy $\left\langle H\right\rangle$ is a monotonic
function of $\beta$ and as a result the intermediates can be
easily chosen (see appendix for details) and the calculation is
much easier to verify. Moreover, the monotonicity of the function
may result in less intermediates and facilitate the calculation
(the number of integration points scales as the free energy
difference which is roughly linear with the number of particles
$N$).

The method was applied to calculate free energy differences in a
toy model of hard rods on a 1-dimensional ring.% It was also
%applied to calculate free energy differences between interior
%loops of RNA that differ in a base (See \cite{AsafThesis} for details).

There is a provisional patent pending that includes the contents of this paper. This work was partially supported by the Leir Charitable Foundation (ED,
AF) the Einstein Center for Theoretical Physics (NC) and the National
Science Foundation under CHE-0713981 (CHM). We want to acknowledge Amir Marcovitz
 for his assistance.
\section*{Appendix: Integration method in the Temperature Integration procedure}

We introduce the method in which the sets of $\beta$-values are chosen
in the process of sampling the function in the TempI procedure. These
sets have to be chosen in a way that will minimize the total error
calculated according to the Simpson's method and will satisfy optimal
acceptance rates in the PT procedures.

We chose a temperature $T_\mathrm{PT}$ -- higher than the energy barriers in the system.
In each PT procedure we kept
constant the number of $\beta$-values sampled within the range $\left[T_{1},T_\mathrm{PT}\right]$,
used in the PT procedure,
and added them to a set of $\beta$-values in the range $\left[T_\mathrm{PT},T_{n}\right]$.

In this method we first sample over the set of $\beta$-values chosen to
satisfy optimal acceptance rates in the system, including the points
$\left(\beta_{n}+\beta_\mathrm{PT}\right)/2$ and $\beta_{n}$,
as defined here and in Sec. \ref{sub:Temperatures-Integration}.

Second, we sample over a set of $\beta$-values that bisect the intervals
of the former set. In the subsequent bisecting we take into account
the facts that in the range $\left[T_{1},T_\mathrm{PT}\right]$ the requirements
for PT go together with the ones needed for optimal sampling of the
integral, and that in the range $\left[T_\mathrm{PT},T_{n}\right]$ the two
requirements don't necessarily correlate but the temperatures can
be sampled independently of the value of the other temperatures in
the set.

The 5 sampling points in the range $\left[T_\mathrm{PT},T_{n}\right]$ we
now have form a subinterval according to ASQ. We further sample this
subinterval according to GASQ until the maximal error in the range
(defined by $\Delta x_{i}\Delta y_{i}$) is smaller than the maximal
error in the range $\left[T_{1},T_\mathrm{PT}\right]$. The group of subintervals
in this range will form a vector which will be called $v_{\mathrm{high\, temps}}$.

We now further bisect the sampling points in the range $\left[T_{1},T_\mathrm{PT}\right]$
in 2 PT procedures, with the subinterval with the maximal error in
$v_{\mathrm{high\, temps}}$ (according to GASQ). We repeat this step,
necessitating this time 4 PT procedures and the sampling of the 2
subintervals with maximal error in $v_{\mathrm{high\, temps}}$.

Then, we generate 2 arrays of vectors of subintervals in the first
range. The first consists of the odd subintervals and the second of
the even subintervals, each vector consisting of the subintervals
that will be derived from the original subinterval. We perform a loop
in which in each iteration, we choose one of the 2 arrays. Then we
select in each vector in the array the subinterval with the maximal
error and we add to the chosen set the subinterval with the maximal
error in $v_{\mathrm{high\, temps}}$. We bisect the chosen set of
subintervals and since each bisecting includes 4 sampling points,
we perform 4 PT procedures in order to do so. Then the bisected subintervals
are placed instead of the subinterval from which they were derived,
and the integral is calculated. Thus, we proceed until a certain total
error or to a maximal number of iterations is reached. In each calculation
of the integral, the number of MC steps performed is registered.

%"CHM is supported by the National Science Foundation under CHE-0713981."

\begin{spacing}{0.5}
\bibliographystyle{unsrt}
{\footnotesize
\bibliography{bib12}}
\end{spacing}
\end{document}